\documentclass[12pt]{iopart}

\usepackage{iopams}
\usepackage{graphicx}
\usepackage{bm}
\usepackage{color}
\eqnobysec

\begin{document}

\title{Sufficient conditions for uniqueness of the weak value}

\author{J. Dressel and A. N. Jordan}
\address{Department of Physics and Astronomy, University of Rochester, Rochester, New York 14627, USA}
\eads{\mailto{jdressel@pas.rochester.edu}, \mailto{jordan@pas.rochester.edu}}

\date{\today}


\def\la{\langle}
\def\ra{\rangle}

\newcommand{\op}[1]{\hat{\bm #1}}                
\newcommand{\mean}[1]{\la#1\ra}                  
\newcommand{\cmean}[2]{_{#1}\mean{#2}}           
\newcommand{\ket}[1]{\vert#1\ra}                 
\newcommand{\bra}[1]{\la#1\vert}                 
\newcommand{\ipr}[2]{\la#1\vert#2\ra}            
\newcommand{\opr}[2]{\ket{#1}\bra{#2}}           
\newcommand{\trace}[1]{\Tr\left(#1\right)}    

\begin{abstract}
  We review and clarify the sufficient conditions for uniquely defining the generalized weak value as the weak limit of a conditioned average using the contextual values formalism introduced in Dressel J, Agarwal S and Jordan A N 2010 \PRL \textbf{104} 240401.  We also respond to criticism of our work in [arXiv:1105.4188v1] concerning a proposed counter-example to the uniqueness of the definition of the generalized weak value.  The counter-example does not satisfy our prescription in the case of an underspecified measurement context.  We show that when the contextual values formalism is properly applied to this example, a natural interpretation of the measurement emerges and the unique definition in the weak limit holds.  We also prove a theorem regarding the uniqueness of the definition under our sufficient conditions for the general case.  Finally, a second proposed counter-example in [arXiv:1105.4188v6] is shown not to satisfy the sufficiency conditions for the provided theorem.
\end{abstract}

\submitto{\JPA}

\maketitle

\section{Introduction}

Since its definition in 1988 by Aharonov, Albert, and Vaidman, the weak value~\cite{AAV} of a quantum operator $\op{A}$ has been a source of considerable controversy.  The formal weak value expression, 
\begin{eqnarray}\label{eq:AAV}
  A_w = \frac{\bra{\psi_f} \op{A} \ket{\psi_i}}{\ipr{\psi_f}{\psi_i}},
\end{eqnarray}
was originally derived as the weak coupling limit of the shift in the mean of a Gaussian momentum pointer $\op{p}$ under a specific von Neumann interaction Hamiltonian $\op{H}_I(t) = -g(t) \op{q}\otimes\op{A}$ that coupled the conjugate position pointer $\op{q}$ to a system observable $\op{A}$ subject to the double boundary conditions of a pure initial preparation state $\ket{\psi_i}$ and a pure final post-selection state $\ket{\psi_f}$.  Though the conditioned pointer shift lent itself to a natural interpretation as a conditioned average, the weak value expression \eref{eq:AAV} violated such intuition by exceeding the eigenvalue range of the observable $\op{A}$ and even being complex.  Despite later experimental confirmation of the effect~\cite{exps}, there was a feeling that such a strange quantity would prove to be an anomalous curiosity.

Far from being an anomaly, however, the formal weak value expression \eref{eq:AAV} has persisted in the literature as a relatively stable quantity in a diverse array of systems.  Ironically, the same features that made its interpretation troublesome have since been been fruitfully used to theoretically address a number of conceptual difficulties in quantum mechanics, including the three-box paradox, Hardy's paradox, superluminal travel, Bohmian trajectories, complementarity, macrorealism violation, and contextuality \cite{theory}.  More practically, the inflation from the eigenvalue range has been exploited to amplify small signals above the background noise, in polarization and interferometric experiments \cite{expamp}.

Given its increasingly common presence in the literature, there was considerable motivation to find a firmer foundation under which the formal expression \eref{eq:AAV} could be understood as a generally measurable feature related to an observable in a pre- and post-selected ensemble.  Recently we provided such a foundation in the form of a Physical Review Letter \cite{Dressel} that indicated how the quantum weak value could be subsumed as an idealized special case of a more flexible operational formalism for the generalized measurement of observables, which we dubbed the \emph{contextual values} formalism.  Our Letter indicated that a principled generalization of the weak value,
\begin{eqnarray}\label{eq:DAJ}
  \cmean{f}{A}_w = \frac{\trace{\op{E}^{(2)}_f (\op{A}\op{\rho} + \op{\rho}\op{A})}}{2\trace{\op{E}^{(2)}_f \op{\rho}}},
\end{eqnarray}
could be uniquely defined as the weak measurement limit of the most general empirical conditioned average \emph{under certain conditions} from a mixed initial state $\op{\rho}$ and an unsharp post-selection represented by an arbitrary probability operator (or POVM element) $\op{E}^{(2)}_f$.  The generalization \eref{eq:DAJ} reduces to the \emph{real part} of \eref{eq:AAV} for pure states, clarifying the origin and significance of the formal expression \eref{eq:AAV} from a broader perspective.  Detailed discussion on the derivation was to be saved for a longer paper reviewing and extending the full theory of contextual values, which has now also been posted \cite{Dressel2}.  

The conditions under which the generalized weak value \eref{eq:DAJ} can be uniquely defined as a limit point of a conditioned average have become recently contested in six versions of a lengthy arXiv paper \cite{ParrottCV} and a summary of the same \cite{ParrottCV2}.  The latter presents concise proposed counter-examples to the uniqueness of the definition \eref{eq:DAJ} based on the understanding of our work, to which we now reply.  The basic issue at hand is quite simple: under what conditions can one obtain the result \eref{eq:DAJ} as the limit point of a conditioned average? 

As we explicitly mention in \cite{Dressel}, the conditioned average does not generally converge to \eref{eq:DAJ} in the weak measurement limit; indeed, the limit can depend on the details of the detection setup, which we call the measurement context.  We stress that our result \eref{eq:DAJ} is thus not in contradiction to the general observation that \emph{the weak value is not a unique limit point of a conditioned average}, which has been previously reported \cite{unique}.  The sole issues being clarified here are the sufficient conditions for obtaining the context-independent special case \eref{eq:DAJ} from the general form of the conditioned average.

This paper is devoted entirely to the subject of the uniqueness of the definition of the generalized weak value and is organized as follows.  In section \ref{sec:cv} we review some elements of the contextual value formalism.  In section \ref{sec:ce} we analyze a proposed counter-example from \cite{ParrottCV2} with the contextual value formalism.  In section \ref{sec:pi} we review the motivation behind our protocol for contextual value assignment.  This is followed in section \ref{sec:proof} by a general theorem and proof of our original definition in \cite{Dressel} along with a precise statement of the sufficient conditions for our theorem to hold.  After discussion of the theorem in section \ref{sec:disc}, we analyze a second proposed counter-example from \cite{ParrottCV3} in section \ref{sec:ce2}.  Finally, we give our conclusions in section \ref{sec:conc}.

\section{Contextual Value formalism}\label{sec:cv}
To keep this work self-contained, we briefly review the contextual values formalism introduced in \cite{Dressel} and expanded upon in \cite{Dressel2}.  The central observation of the contextual values formalism is that an observable $\op{A}$ for a particular system can be completely measured indirectly using an imperfectly correlated detector.  The formalism is powerful enough to subsume strong measurements, weak measurements, and any strength of measurement in between.  Indeed, the von Neumann measurement used to derive the weak value \eref{eq:AAV} originally becomes a special case.

For the typical case of a detector with a pure preparation state $\ket{d}$ that is coupled to the system with any joint unitary operation $\op{U}_{sd}$ and then subsequently measured in a detector basis $\{\ket{j}\}$, such an indirect measurement will be completely characterized by a set of measurement operators on the system $\{\op{M}_j = \bra{j}\op{U}_{sd}\ket{d}\}$, which we call a \emph{measurement context}.  As in \cite{Dressel}, we restrict ourselves to this typical case in what follows for simplicity; the straight-forward generalization to impure detector preparation is detailed in \cite{Dressel2}.

When a system state $\op{\rho}$ is conditioned on a particular outcome $j$ of the detector, it becomes updated according to $\op{\rho}_j = \op{M}_j \op{\rho} \op{M}_j^\dagger / P(j)$, where the normalization probability for detecting the outcome $j$ is given by $P(j) = \trace{\op{\rho} \op{E}_j}$.  The positive probability operators $\{\op{E}_j = \op{M}_j^\dagger \op{M}_j\}$ partition unity $\sum_j \op{E}_j = \op{1}$, forming a positive operator-valued measure (POVM) on the system space.

The expectation value of the observable $\op{A}$ can be accurately measured by the imperfectly correlated detector provided that the following operator identity exists,
\begin{eqnarray}
  \label{eq:cv}
  \op{A} = \sum_j \alpha_j \op{E}_j, \\
  \mean{A} = \trace{\op{\rho} \op{A}} = \sum_j \alpha_j P(j),
\end{eqnarray}
which defines the \emph{contextual values} $\{\alpha_j\}$ of the observable $\op{A}$ with respect to the measurement context $\{\op{M}_j\}$.  As we shall explain in section \ref{sec:pi}, in the event that multiple solutions for the contextual values exist we prescribe picking the solution that places the tightest bound on the detector variance, which can be found using the pseudoinverse.

If the observable $\op{A}$ also commutes with the entire measurement context $\forall j,\, [\op{A},\op{M}_j]=0$, then all the statistical moments of $\op{A}$ can also be accurately measured by correlating \emph{sequences} of measurements on the detector,
\begin{eqnarray}
  \label{eq:moments}
  \mean{A^n} = \sum_{j_1 \ldots j_n} (\alpha_{j_1}\cdots\alpha_{j_n}) \trace{\op{\rho}\op{E}_{j_1}\cdots\op{E}_{j_n}}.
\end{eqnarray}
We call a detector that can measure all moments of $\op{A}$ a \emph{fully compatible} detector.  In what follows we will concern ourselves mostly with fully compatible detectors.

For the special case of a projective detector, the measurement context $\{\op{\Pi}_k\}$ consists of the spectral projections of $\op{A}$, so \eref{eq:cv} reduces to the spectral expansion $\op{A} = \sum_k a_k \op{\Pi}_k$ as a special case, where $a_k$ are the eigenvalues of $\op{A}$, and \eref{eq:moments} reduces to the standard formula $\mean{A^n} = \sum_k a_k^n P(k)$ that needs only a single repeated measurement to obtain all moments.  Hence, the contextual values can be considered to form a generalized spectrum for the observable that is specific to a particular measurement context.

If a second measurement is made after the first measurement of $\op{A}$ that is characterized by an arbitrary second measurement context and associated probability operators $\{\op{E}^{(2)}_f\}$, we can also construct the most general \emph{conditioned averages} of the observable,
\begin{eqnarray}
  \label{eq:condav}
  \cmean{f}{A} = \sum_j \alpha_j P(j | f), \\
  P(j | f) = \frac{\trace{\op{E}^{(2)}_f \op{M}_j\op{\rho}\op{M}_j^\dagger}}{\sum_j \trace{\op{E}^{(2)}_f \op{M}_j\op{\rho}\op{M}_j^\dagger}}.
\end{eqnarray}
The post-selected conditional probabilities $P(j | f)$ are generalizations of the Aharonov-Bergmann-Lebowitz rule \cite{theory} that handle mixed states, general intermediate measurement, and unsharp post-selections.  As the conditioned averages \eref{eq:condav} are constructed entirely from measurable quantities, they form a principled foundation for deriving the generalization of the weak value \eref{eq:DAJ} as a limiting value as the correlation between the system and detector vanishes.

In what follows, we stress that the contextual values formalism itself has not been challenged.  Only the details of the derivation of the context-independent weak value \eref{eq:DAJ} using the general conditioned average \eref{eq:condav} are being contested.

\section{Analysis of a counter-example}\label{sec:ce}

We now address the counter-example provided in \cite{ParrottCV2}.  A case where the number of POVM elements (or measurement operators in this case) exceeds the dimension of the Hilbert space for a system observable $\op{A}$ is considered therein, 
\begin{eqnarray}\label{eq:pmeas}
  \eqalign{
  \op{M}_1 = \left(\begin{array}{cc}1/2 + g & 0 \\ 0 & 1/2 - g\end{array}\right), \\
  \op{M}_2 = \left(\begin{array}{cc}1/2 - g & 0 \\ 0 & 1/2 + g\end{array}\right), \\
  \op{M}_3 = \sqrt{1/2 - 2g^2}\;  \op{1}, } \\
  \op{A} = \left(\begin{array}{cc}a & 0 \\ 0 & b\end{array}\right),
\end{eqnarray}
where the operators are expressed as matrices in the basis that diagonalizes $\op A$.  

To calibrate the measurement, one is then faced with determining contextual values $\{\alpha_j\}$ that satisfy the (now underspecified) equation \eref{eq:cv}.  To see this in detail, since all operators commute and are diagonalized in the same basis, we can write \eref{eq:cv} as the equivalent matrix equation, $\vec{a} = F \vec{\alpha}$, where $F_{kj} = \trace{\op{\Pi}_k \op{E}_j}$:
\begin{eqnarray}
  \left(\begin{array}{c}a\\b\end{array}\right) = \left(\begin{array}{ccc}(1/2 + g)^2 & (1/2 - g)^2 & 1/2 - 2g^2 \\ (1/2 - g)^2 & (1/2 + g)^2 & 1/2 - 2g^2\end{array}\right)\left(\begin{array}{c}\alpha_1 \\ \alpha_2 \\ \alpha_3\end{array}\right).
\end{eqnarray}

This underspecified matrix equation is then solved in \cite{ParrottCV2} by choosing $\alpha_1 = 1/g^2$ arbitrarily and then solving the resulting modified equation,
\begin{eqnarray}
  \left(\begin{array}{c}a - \frac{(1/2 + g)^2}{g^2} \\b - \frac{(1/2 - g)^2}{g^2}\end{array}\right) = \left(\begin{array}{cc}(1/2 - g)^2 & 1/2 - 2g^2 \\ (1/2 + g)^2 & 1/2 - 2g^2\end{array}\right)\left(\begin{array}{c}\alpha_2 \\ \alpha_3\end{array}\right),
\end{eqnarray}
which leads to the full solution,
\begin{eqnarray}\label{eq:ppsol}
  \eqalign{
  \alpha_1 &= \frac{1}{g^2}, \\
  \alpha_2 &= \frac{1}{g^2} - \frac{a - b}{2g}, \\
  \alpha_3 &= \frac{4 - g(a(1+2g)^2 - b(1-2g)^2 - 16g)}{4g^2(4g^2-1)}, \\
  &= -\frac{1}{g^2} + \frac{a - b}{4g} + (a + b - 8) + O(g), }
\end{eqnarray}
which contains poles of order $1/g^2$ by construction.  These poles then contribute an extra context-dependent term to the weak limit of \eref{eq:condav} that is not included in \eref{eq:DAJ}.  For the specific choice of the identity $a=b=1$ considered in \cite{ParrottCV2}, then $\alpha_1 = \alpha_2 = 1/g^2$ and $\alpha_3 = (2g^2 + 1)/(g^2(4g^2-1))$.

We devote a considerable amount of space to this type of underspecified case in our four page Letter \cite{Dressel}.  We write on page 2, ``The latter case [where the number of POVM elements exceeds the dimension of the system operator] results in an infinite number of possible solutions, ${\alpha_j}$.  As such, we propose that the physically sensible choice of [contextual values] is the least redundant set uniquely related to the eigenvalues through the Moore-Penrose pseudoinverse.''  All examples we give in the paper use the pseudoinverse, and this discussion occurs immediately before the \emph{conditioned average} section under contention.

The problem with the counter-example is that the pseudoinverse solution is not employed, and consequently the freedom in the set of underspecified equations is used to insert by hand an anomalous contextual value that diverges as $1/g^2$ in order to artificially produce an extra contribution to the result \eref{eq:DAJ} in the $g\to0$ weak limit.  Indeed, we could go further by similarly choosing a contextual value that diverges as $g^{-m}$, where $m>2$.  Such a case would produce a formally divergent conditioned average in the weak limit.  

However, if we solve for the contextual values using the prescription we describe in our paper, the assignment gives a clear physical interpretation to the measurement that is being done.  The pseudoinverse solution is found from the singular value decomposition, $F = U \Sigma V^T$.  For this example, we find,
\begin{eqnarray}
  \eqalign{
  U = \frac{1}{\sqrt{2}}\left(\begin{array}{cc}-1 & 1 \\ 1 & 1\end{array}\right), \\
  V = \frac{1}{\sqrt{2}}\left(\begin{array}{ccc}-1 & \frac{4g^2 + 1}{\sqrt{48 g^4 - 8 g^2 + 3}} & \frac{\sqrt{2}(4g^2 - 1)}{\sqrt{48 g^4 - 8 g^2 + 3}} \\ 1 & \frac{4g^2 + 1}{\sqrt{48 g^4 - 8 g^2 + 3}} & \frac{\sqrt{2}(4g^2 - 1)}{\sqrt{48 g^4 - 8 g^2 + 3}} \\ 0 & \frac{-2(4g^2 - 1)}{\sqrt{48 g^4 - 8 g^2 + 3}} & \frac{\sqrt{2}(4g^2 + 1)}{\sqrt{48 g^4 - 8 g^2 + 3}}\end{array}\right), \\
  \Sigma = \left(\begin{array}{ccc}2g & 0 & 0 \\ 0 & \frac{1}{2}\sqrt{48 g^4 - 8 g^2 + 3} & 0\end{array}\right). }
\end{eqnarray}
The pseudoinverse is then $F^+ = V \Sigma^+ U^T$, where $\Sigma^+$ is a diagonal matrix inverting all nonzero elements of $\Sigma^T$.  We can then find our prescribed solution as ${\vec \alpha} = F^+ {\vec a}$, which for this example is,
\begin{eqnarray}\label{eq:psol}
  \eqalign{
  \alpha_1 = \frac{a-b}{4g} + \frac{(a+b)(4g^2 + 1)}{48g^4 - 8g^2 + 3} = \frac{a-b}{4g} + \frac{a+b}{3} + O(g^2), \\
  \alpha_2 = -\frac{a-b}{4g} + \frac{(a+b)(4g^2 + 1)}{48g^4 - 8g^2 + 3} = -\frac{a-b}{4g} + \frac{a+b}{3} + O(g^2), \\
  \alpha_3 = \frac{2(a+b)(1 - 4g^2)}{48g^4 - 8g^2 + 3} = \frac{2(a+b)}{3} + O(g^2). }
\end{eqnarray}

The largest pole in the solution \eref{eq:psol} has order $1/g$, which is the inverse of the smallest nonzero order of $g$ in the POVM generated by \eref{eq:pmeas}---we will show this is the general rule for pseudoinverse solutions that correctly satisfy $\vec{a} = F\vec{\alpha}$ with the lowest nonzero order in $g$.  It is then easy to check that the generalized weak value \eref{eq:DAJ} will be recovered from the conditioned average \eref{eq:condav} in the weak limit as $g\to 0$ for any pre- and post-selection, as claimed.  

For the special case of the identity, $a=b=1$, that is considered, the solution \eref{eq:psol} does not diverge as $g\to0$, but actually converges to a constant.  This behavior is intuitive because the measured system operator is the identity---the identity can always be constructed from the $g=0$ POVM alone.  In this case, the first two contextual values converge to the same value of $2/3$, while the third contextual value converges to $4/3$ and contributes twice as much to the average; this makes physical sense as the first two outcomes balance each other to produce the identity, while the third outcome directly corresponds to the identity being measured.  Moreover, for the orthogonal case $a=1,b=-1$ the first two contextual values simplify to $\pm(1/2g)$, while the third contextual value vanishes entirely; this makes physical sense since the third outcome is orthogonal to the operator being measured and can therefore be discarded.  None of these physically intuitive features are present in the solution \eref{eq:ppsol} presented in \cite{ParrottCV2}.

\section{Pseudoinverse prescription}\label{sec:pi}
It is now worthwhile to review the pseudoinverse prescription, and to discuss its methodology and advantages.  We recall that the equation we are solving is \eref{eq:cv} in the form of the matrix equation $\vec{a} = F \vec{\alpha}$, where $F$ is an $N\times M$ matrix ($N$ being the dimension of the system, and $M$ being the number of POVM elements) given by its elements, $F_{kj} = \trace{\op{\Pi}_k \op{E}_j}$.  We can then decompose this matrix with the singular value decomposition, $F = U \Sigma V^T$, where $U$ is an $N \times N$ orthogonal matrix, $V$ is an $M \times M$ orthogonal matrix, and $\Sigma$ is a $N \times M$ diagonal matrix of singular values.  The pseudoinverse of $F$ is then constructed as $F^+ = V \Sigma^+ U^T$, where $\Sigma^+$ is a $M \times N$ diagonal matrix formed by inverting the non-zero singular values.  The pseudoinverse reduces correctly to the true inverse if one exists.

With the pseudoinverse in hand, we then find a uniquely specified solution $\vec{\alpha}_0 = F^+ \vec{a}$ that is directly related to the eigenvalues of the operator.   Other solutions $\vec{\alpha} = \vec{\alpha}_0 + \vec{x}$ of \eref{eq:cv} will contain additional components in the null space of $F$, and will thus deviate from this least redundant solution.  Consequently, the solution $\vec{\alpha}_0$ has the least norm of all solutions, since $|| \vec{\alpha} ||^2 = || \vec{\alpha}_0 ||^2 + || \vec{x} ||^2$ by the triangle inequality and the fact that $\vec{\alpha}_0$ and $\vec{x}$ live in orthogonal subspaces.  Even in the case of an overdetermined set of equations (where the number of detector outcomes is less than the dimension of the system), the pseudoinverse will give the ``best fit'' solution in the least-squares sense.   This can be seen by solving $F^T F \vec{\alpha} = F^T \vec{a}$.  One will also obtain $\vec{\alpha} = \vec{\alpha}_0 + \vec{x}$, where now $\vec{\alpha}_0$ does not solve $F \vec{\alpha} = \vec{a}$, but is the least squares fit to it, and $\vec{x}$ is in the null space of $F^T F$.  As a physical example of this last situation one could use a grid of point measurements like a pixel array to approximate measurements for a continuous variable, such as position.

In addition to the mathematical reasons for using the pseudoinverse in this context, there is an important physical one that we will now describe.  As mentioned, a fully compatible detector can be used together with the contextual values to reconstruct any moment of a compatible observable.  However, since the detector outcomes are imperfectly correlated with the observable, the contextual values typically lie outside of the eigenvalue range and many repetitions of the measurement must be practically performed to obtain adequate precision for the moments.  Importantly, the uncertainty in the moments is controlled by the variance---not of the observable operator, but of the contextual values themselves,
\begin{eqnarray}
  \sigma^2  = \sum_j \alpha_j^2 P(j) - \mean{A}^2,
\end{eqnarray}
where $P(j)$ is the probability of outcome $j$.   Since the mean of the contextual values is set by construction to the mean of the observable being measured, it is in the experimentalist's best interest to minimize the second moment of the contextual values.  This moment has a simple upper bound of $\sum_j \alpha_j^2 P(j) < \sum_j \alpha_j^2 = || {\vec \alpha}||^2$ because $0 < P(j) < 1$, which will also constitute an upper bound of the variance $\sigma^2$.  In absence of prior knowledge about the system one is dealing with, this is a reasonable upper bound to make.  Therefore, by minimizing this upper bound the pseudoinverse will choose a solution that provides rapid statistical convergence for observable measurements on the system given no prior knowledge of the system state.

For the case of the counterexample in \cite{ParrottCV2}, the solution \eref{eq:ppsol} has to leading order the bound on the variance,
\begin{eqnarray}
  ||\vec{\alpha}||^2 = \frac{3}{g^4} - \frac{3(a-b)}{2 g^3} + O\left(\frac{1}{g^2}\right),
\end{eqnarray}
while the pseudoinverse solution \eref{eq:psol} has to leading order the bound,
\begin{eqnarray}
  ||\vec{\alpha}||^2 = \frac{(a-b)^2}{8 g^2} + \frac{2}{3}(a+b)^2 + O(g^2).
\end{eqnarray}
For any observable $\vec{a}$ the solution \eref{eq:ppsol} has a detector variance bounded by leading order $1/g^4$, which could generally swamp any attempt to measure an observable near the weak limit.  In particular, the conditioned averages \eref{eq:condav} would not generally be tractable to obtain, so the anomalous weak limit derived in \cite{ParrottCV2} may not be easily observable without a special initial state.  However, the pseudoinverse solution \eref{eq:psol} has a detector variance bounded by leading order $1/g^2$ in the worst case; moreover, for the identity, $a=b=1$, then the bound on the noise minimizes to a constant as $g\to 0$.

\section{General theorem}\label{sec:proof}
We now give a general proof of the result \eref{eq:DAJ}.  To obtain this result, we make the following sufficient assumptions:
\begin{enumerate}
  \item The measurement operators $\{\op{M}_j\}$ are analytic functions of a measurement strength parameter $g$, and thus have well defined Taylor expansions around $g=0$ such that $\forall j, \, \lim_{g\to 0}\op{M}_j \propto \op{1}$.  This is physically reasonable because measurement operators are typically composed from matrix elements of an analytic evolution operator under an interaction Hamiltonian for which $g$ is the coupling constant.
  \item If a measurement operator $\op{M}_j = \op{U}_j \op{E}_j^{1/2}$ is not positive, its unitary freedom $\op{U}_j = \exp(i g \op{G}_j)$ is generated by a Hermitian operator $\op{G}_j$ that commutes with the density matrix $\op{\rho}$ of the system, $\forall j,\, [\op{G}_j,\op{\rho}] = 0$.  The reason for this assumption will become clear.
  \item The equality $\op{A} = \sum_j \alpha_j(g) \op{E}_j(g)$ must be satisfied, where the contextual values $\alpha_j(g)$ are selected according to the pseudoinverse prescription.
  \item The minimum nonzero order in $g$ for all $\op{E}_j(g)$ is $g^n$ such that (iii) is satisfied.  (In \cite{Dressel} we considered the typical case $n=1$.) 
  \item The POVM elements $\{\op{E}_j\}$ all commute with the observable $\op{A}$, so that they are diagonalizable in the same basis.
\end{enumerate}
Then we have the following theorem: in the weak limit $g\to0$ the context dependence of the conditioned average \eref{eq:condav} vanishes and the generalized weak value \eref{eq:DAJ} is uniquely defined.

We note before we prove this result that these are only the sufficient conditions for the unique definition \eref{eq:DAJ} that we implied in \cite{Dressel}---some of the assumptions might be further weakened.  For example, there may be other principled inversion schemes for the contextual values that also lead to the context-independent result \eref{eq:DAJ}.

To obtain the proof, we shall rewrite \eref{eq:condav} in a useful form and then take the weak limit as $g\to 0$.  Using the polar decomposition of the measurement operators $\op{M}_j = \op{U}_j\op{E}^{1/2}_j$, we rewrite the probabilities that appear in \eref{eq:condav} as,
\begin{eqnarray}
  \label{eq:proof1}
  \trace{\op{E}^{(2)}_f\op{M}_j\op{\rho}\op{M}^\dagger_j} = \trace{(\op{U}_j^\dagger \op{E}^{(2)}_f \op{U}_j) \op{\rho}'_j},
\end{eqnarray}
where the modified density operator is,
\begin{eqnarray}
  \label{eq:proof2}
  \op{\rho}'_j = \op{E}^{1/2}_j\op{\rho}\op{E}^{1/2}_j = \frac{1}{2}\{\op{E}_j,\op{\rho}\} - \frac{1}{2}[\op{E}^{1/2}_j,[\op{E}^{1/2}_j,\op{\rho}]],
\end{eqnarray}
and $\{a,b\} = ab + ba$ denotes the anticommutator.

From assumptions (i) and (iv) we have the lowest nonzero order expansion of the POVM $\op{E}_j = p_j \op{1} + g^n \op{E}^{(n)}_j + O(g^{n+1})$ where $p_j \in (0,1)$ are nonzero probabilities such that $\sum_j p_j = 1$.  We therefore also have the expansion of the positive roots to the same order in $g$,
\begin{eqnarray}\label{eq:weakexpansions}
  \op{E}^{1/2}_j(g) = \sqrt{p_j}\op{1} + g^n \op{E}^{(n)}_j/2\sqrt{p_j} + O(g^{n+1}).
\end{eqnarray}
The probabilities $p_j$ must be nonzero to satisfy assumption (i).  The physical probability of outcome $j$ is given by $P(j) = \trace{\rho \op{E}_j}$, and therefore converges to $p_j$ in the weak limit, $g \rightarrow 0$.  

Inserting the expression \eref{eq:weakexpansions} into \eref{eq:proof2}, we find
\begin{eqnarray}  \label{eq:proof3}
 \op{\rho}'_j = p_j \op{\rho} + \frac{g^n}{2}\{\op{E}^{(n)}_j,\op{\rho}\} - \frac{g^{2n}}{8 p_j}[\op{E}^{(n)}_j,  [\op{E}^{(n)}_j ,\op{\rho}]].
\end{eqnarray}
This leaves the probabilities that appear in \eref{eq:condav} to be,
\begin{eqnarray}
 \label{eq:proof4} \fl
 \trace{\op{E}^{(2)}_f\op{M}_j\op{\rho}\op{M}^\dagger_j} = 
  p_j \trace{(\op{U}_j^\dagger \op{E}^{(2)}_f \op{U}_j) \op{\rho}} + \frac{g^n}{2} \trace{(\op{U}_j^\dagger \op{E}^{(2)}_f \op{U}_j) \{\op{E}^{(n)}_j,\op{\rho}\}},
\end{eqnarray}
plus a correction of order $O(g^{2n})$.  

Invoking assumption (ii), since the generators $\op{G}_j$ of unitaries commute with the density matrix, the unitary itself commutes with the density matrix.  Consequently, the first term in the righthand side of \eref{eq:proof4} simplifies to   $p_j \trace{\op{E}^{(2)}_f  \op{\rho}}$.  In the term of order $O(g^n)$, we can expand the unitary operator to first order in $g$, $U_j = \op{1} + i g \op{G}_j + O(g^2)$ to find that the second term in the righthand side of \eref{eq:proof4} simplifies to 
$(g^n/2) \trace{ \op{E}^{(2)}_f \{\op{E}^{(n)}_j,\op{\rho}\}}$ plus a correction of order $O(g^{n+1})$.  

Thus, we find that the denominator of \eref{eq:condav} is
\begin{eqnarray}\label{eq:denominator}
 \sum_j p_j \trace{\op{E}^{(2)}_f  \op{\rho}} + \sum_j (g^n/2) \trace{ \op{E}^{(2)}_f \{\op{E}^{(n)}_j,\op{\rho}\}} + O(g^{n+1}).
\end{eqnarray}
However, since $\sum_j p_j = 1$ and $\sum_j \op{E}^{(n)}_j =0$ (the POVM condition), the denominator is simply $\trace{\op{E}^{(2)}_f  \op{\rho}}$, with a correction of order $O(g^{n+1})$.

The numerator of \eref{eq:condav} is given by summing \eref{eq:proof4} with the contextual values to find,
\begin{eqnarray}\label{eq:numerator}
\sum_j \alpha_j  \trace{\op{E}^{(2)}_f (\frac{1}{2} \{ p_j \op{1} + g^n \op{E}^{(n)}_j, \op{\rho} \})}  + O(g^{n+1}).
\end{eqnarray}
We note that to order $g^n$, $\op{A} = \sum_j \alpha_j(g) (p_j\op{1} + g^n\op{E}^{(n)}_j)$; since this sum exactly appears in the numerator, we recover our original result \eref{eq:DAJ}, up to a numerator correction of order $O(g^{n+1})$ times the order of each $\alpha_j$.  Thus, the only way the result \eref{eq:DAJ} can be spoiled under our assumptions is if $\alpha_j(g)$ has a pole larger than $O(1/g^{n})$.  Hence, the last step in the proof will be to show that the pseudoinverse solution of $\alpha_j(g)$ cannot have a pole larger than $O(1/g^n)$.  

To address the order of the contextual values $\alpha_j(g)$, we will first simplify notation by noting that $\op{A}$ commutes with $\{\op{E}_j(g)\}$ according to assumption (v).  As such, we will replace all the diagonal matrices with vectors and rewrite the contextual values definition \eref{eq:cv} from assumption (iii) as an equivalent matrix equation,
\begin{eqnarray}
  \label{eq:proofcv}
  \vec{a} = F \vec{\alpha},
\end{eqnarray}
where,
\begin{eqnarray}
  \label{eq:F}
  F = \left(\begin{array}{ccc}\vec{E}_1(g) & \vec{E}_2(g) & \dots\end{array}\right) = P + g^n F_n + O(g^{n+1}), 
\end{eqnarray}
and the two leading order matrices are defined as,
\begin{eqnarray}
  \eqalign{
  P = \left(\begin{array}{ccc}p_1 \vec{1} & p_2 \vec{1} & \dots\end{array}\right), \\
  F_n = \left(\begin{array}{ccc}\vec{E}_1^{(n)} & \vec{E}_2^{(n)} & \dots\end{array}\right). }
\end{eqnarray}

As discussed, the minimum norm solution to \eref{eq:proofcv} is the pseudoinverse solution $\vec{\alpha}_0 = F^+ \vec{a}$.  The pseudoinverse is constructed from the singular value decomposition $F = U\Sigma V^T$ as $F^+ = V\Sigma^+ U^T$, where $U$ and $V$ are orthogonal matrices such that $U^T U = V V^T = 1$, $\Sigma$ is the singular value matrix composed of the square roots of the eigenvalues of $FF^T$, and $\Sigma^+ $ is composed of the inverse nonzero elements in $\Sigma^T$.  In order to satisfy \eref{eq:proofcv}, then we have the equivalent condition for each component of $U^T\vec{a} = \Sigma V^T \vec{\alpha}$,
\begin{eqnarray}\label{eq:svcond}
  (U^T \vec{a})_k = \Sigma_{kk} (V^T \vec{\alpha})_k.
\end{eqnarray}
Therefore, all singular values $\Sigma_{kk}$ corresponding to nonzero components of $U^T \vec{a}$ must also be nonzero; for brevity we shall call these the relevant singular values.  Singular values which are not relevant will not contribute to the solution $\vec{\alpha} = V \Sigma^+ U^T \vec{a}$.  Since $\alpha_j = (V \Sigma^+ U^T \vec{a})_j = \sum_k V_{jk} \Sigma^+_{kk} (U^T \vec{a})_k$, any zero element of $U^T\vec{a}$ will eliminate the inverse irrelevant singular value $\Sigma^+_{kk}$ from the solution for $\alpha_j$.  

Since the orthogonal matrices $U$ and $V$ have nonzero orthogonal limits $\lim_{g\to 0} U = U_0$ and $\lim_{g\to 0} V = V_0$, such that $U_0^T U_0 = V_0 V_0^T = 1$, and since $\vec{a}$ is $g$-independent, then the only poles in the solution $\vec{\alpha}_0 = F^+ \vec{a} = V \Sigma^+  U^T \vec{a}$ must come from the inverses of the relevant singular values in $\Sigma^+$.  If a singular value $\Sigma_{kk} = O(g^m)$, then $\Sigma^+_{kk} = 1/\Sigma_{kk} = O(1/g^m)$; therefore, to have a pole of order higher than $O(1/g^n)$ then there must be at least one relevant singular value with a leading order greater than $g^n$.  However, if that were the case then the expansion of $F$ to order $g^n$ would have a relevant singular value of zero and therefore could not satisfy \eref{eq:svcond}, contradicting the assumption (iv) about the minimum nonzero order of the POVM.  Therefore, the pseudoinverse solution $\vec{\alpha}_0 = F^+\vec{a}$ can have no pole with order higher than $O(1/g^n)$ and the theorem is proved.  

\section{Discussion}\label{sec:disc}

As we stated in our Letter \cite{Dressel}, ``we find that as $g\to0$, the weak limit [of \eref{eq:condav}] generally depends explicitly on $\{\op{G}_j\}$ and $\{\alpha_j\}$, and thus will change depending on how it is measured and how the [contextual values] are chosen.''  These dependences are apparent in the proof from equations \eref{eq:proof1} through \eref{eq:numerator}.  In other words, we find that the weak limit of the conditioned average \emph{is not generally unique}.  However, to produce any limiting value other than \eref{eq:DAJ} one needs to violate the sufficient conditions given for our theorem.  Namely, to find a different weak limit one needs either a nonanalytic or incompatible measurement context, a unitary disturbance that persists in the weak limit, a minimum nonzero order of $g$ that does not satisfy the observable identity, or pathologically chosen CV.

In \eref{eq:proof2} the positive root of the POVM element $\op{E}_j$ performs the information extraction of the measurement and modifies $\op{\rho}$ to $\op{\rho}'_j$, which consists of two terms: a symmetric term involving the POVM element itself, and a double-commutator involving the roots.  The symmetric term leads to the weak value \eref{eq:DAJ}, while the commutator term produces measurement disturbance away from \eref{eq:DAJ}.  The unitary part of the POVM element in \eref{eq:proof1} rotates the post-selection $\op{E}^{(2)}_f$ to a different post-selection that depends explicitly on the measurement result obtained, so this also disturbs the measurement process independently of the information extraction of the measurement.

For this reason, we consider a measurement consisting solely of positive POVM roots to be a \emph{minimally disturbing} measurement, which is consistent with the usage of the term by Wiseman and Milburn \cite{Wiseman}.  That is, the information extraction of the measurement necessarily disturbs the system state by a minimum amount, but no additional unitary rotation occurs.  Note that a weak measurement is an independent concept from a minimally disturbing measurement.  

In \cite{Dressel} we named the limit as $g\to0$ under the sufficiency condition $\forall j, \, [\op{G}_j, \op{\rho}] = 0$ placed on the unitary generators $\op{G}_j$ the \emph{minimal disturbance limit} since the measurement operators act like minimally disturbing POVM roots in that limit.  The minimal disturbance definition $\op{U}_j = \op{1}$ becomes a special case.  

\section{Analysis of a second counter-example}\label{sec:ce2}
Shortly after a preprint for this paper was posted, the reference \cite{ParrottCV2} was updated to a sixth version \cite{ParrottCV3} that adds a second proposed counter-example to our theorem, which we now address.  The second proposed counter-example uses a three-outcome POVM to measure an observable in a three-dimensional Hilbert space to avoid any ambiguity related to the contextual values being underspecified.  Specifically, the following measurement operators and observable are employed,
\begin{eqnarray}\label{eq:mce2}
  \eqalign{
  \op{M}_1 = \left(\begin{array}{ccc}\sqrt{1/2 + g} & 0 & 0 \\ 0 & \sqrt{1/2} & 0 \\ 0 & 0 & \sqrt{1/2 + g}\end{array}\right), \\
  \op{M}_2 = \left(\begin{array}{ccc}\sqrt{1/3 + g^2} & 0 & 0 \\ 0 & \sqrt{1/3 + g} & 0 \\ 0 & 0 & \sqrt{1/3}\end{array}\right), \\
  \op{M}_3 = \left(\begin{array}{ccc}\sqrt{1/6 - g - g^2} & 0 & 0 \\ 0 & \sqrt{1/6 - g} & 0 \\ 0 & 0 & \sqrt{1/6 - g}\end{array}\right),} \\
  \op{A}_p = \left(\begin{array}{ccc}1 & 0 & 0 \\ 0 & 0 & 0 \\ 0 & 0 & 0\end{array}\right),
\end{eqnarray}
where we have corrected a minor typo in the definition of $\op{M}_2$ \footnote{The missing square root over the $1/3$ that is needed to satisfy the POVM condition and obtain the contextual values \eref{eq:ce2cv} has been restored.}.  Computing the contextual values required to satisfy the relation $\op{A}_p = \sum_i \alpha_i \op{M}_i^2$ produces,
\begin{eqnarray}\label{eq:ce2cv}
  \eqalign{
  \alpha_1 = \frac{1}{6g^2} - \frac{1}{g}, \\
  \alpha_2 = \frac{1}{6g^2} - \frac{1}{g}, \\
  \alpha_3 = -\frac{5}{6g^2} - \frac{1}{g}. }
\end{eqnarray}
The $1/g^2$ dependence of the contextual values can lead to the conditioned average \eref{eq:condav} having additional context-dependent terms beyond the weak value \eref{eq:DAJ} that are relevant in the weak limit, which seemingly contradicts our theorem.

This example, however, violates sufficiency condition (iv) for our theorem.  Specifically, to first order in $g$---which is the lowest nonzero order---the POVM elements are,
\begin{eqnarray}
  \eqalign{
  \op{E}'_1 = \left(\begin{array}{ccc}1/2 + g & 0 & 0 \\ 0 & 1/2 & 0 \\ 0 & 0 & 1/2 + g\end{array}\right), \\
  \op{E}'_2 = \left(\begin{array}{ccc}1/3 & 0 & 0 \\ 0 & 1/3 + g & 0 \\ 0 & 0 & 1/3\end{array}\right), \\
  \op{E}'_3 = (1/6 - g) \op{1}. }
\end{eqnarray}
While these first order POVM elements do satisfy the POVM condition $\op{E}'_1 + \op{E}'_2 + \op{E}'_3 = \op{1}$, there is no exact solution to the required identity $\op{A}_p = \sum_i \alpha_i \op{E}'_i$.  

We can see this fact by first noting that there is no solution for an arbitrary observable $\op{A}$.  Specifically, if we write the required identity as a matrix equation with $\vec{a} = F \vec{\alpha}$, with,
\begin{eqnarray}
  F = \left(\begin{array}{ccc}1/2 + g & 1/3 & 1/6 - g \\ 1/2 & 1/3 + g & 1/6 - g \\ 1/2 + g & 1/3 & 1/6 - g\end{array}\right),
\end{eqnarray}
then $F^{-1}$ does not exist since $\det(F) = 0$, so there is no general solution $\vec{\alpha} = F^{-1}\vec{a}$.  However, there may still exist specific observables $\vec{a}'$ for which $\vec{a}' = F \vec{\alpha}$ is an underspecified system of equations with an infinite number of valid contextual value solutions.  To rule out such a case for the specific observable $\vec{a}_p = (1,0,0)$, we compute the pseudoinverse solution $\vec{\alpha} = F^+ \vec{a}_p$,
\begin{eqnarray}
  F^+ \vec{a}_p = \left(\begin{array}{c} \frac{72 g^2 + 30 g + 11}{18 g (12 g^2 + 4 g + 3)} \\ - \frac{18 g^2 + 3 g + 8}{9 g (12 g^2 + 4 g + 3)} \\
    - \frac{(6g - 1)^2}{18 g (12 g^2 + 4 g + 3)}\end{array}\right),
\end{eqnarray}
and subsequently compute,
\begin{eqnarray}
  F (F^+ \vec{a}_p) = \frac{1}{2}\left(\begin{array}{c}1 \\ 0 \\ 1\end{array}\right).
\end{eqnarray}
Since this does not equal $\vec{a}_p = (1,0,0)$, then $\vec{a}_p$ is partially in the nullspace of $F^+$ and there can be no exact solution to the required identity $\op{A}_p = \sum_i \alpha_i \op{E}'_i$.

Therefore, sufficiency condition (iv) for our theorem is violated and we do not expect the theorem to hold.  Intuitively, the measurement \eref{eq:mce2} is not sufficiently correlated with the specific observable $\op{A}_p$ as $g\to 0$ to guarantee the weak value \eref{eq:DAJ} as the limit point of the conditioned average \eref{eq:condav}.

Moreover, if another observable $\op{A}$ could be found such that $\op{A} = \sum_i \alpha_i \op{E}'_i$ were satisfiable to first order in $g$ by the pseudoinverse solution, then the discussion after \eref{eq:svcond} in the proof of our theorem would apply.  Hence, higher order poles would not appear in the contextual values, and the generalized weak value \eref{eq:DAJ} would be obtained as the unique limit point of the conditioned average \eref{eq:condav}.

\section{Conclusion}\label{sec:conc}

We have expanded upon and defended the claim made in our Letter \cite{Dressel} that the context-independent generalized weak value \eref{eq:DAJ} can be uniquely defined as a limit point of the conditioned average \eref{eq:condav}, and have given sufficient mathematical assumptions required for the definiton to hold.  Conceptually, the measurement context should depend on a measurement strength parameter $g$ such that it reduces to the identity as $g\to 0$; any additional unitary disturbance in the measurement should not affect the state above and beyond the measurement being performed; the observable should be measurable to the lowest nonzero order in $g$; the contextual values of the measurement should be chosen to minimize an upper bound for the detector variance; and, the probability operators for the measurement should commute with the observable.  

We have also addressed two counter-examples to our definition that were proposed in versions 1 through 6 \footnote{Version 6 is the most recent version available at the time of writing, so any versions that subsequently appear are outside the scope of this paper.} of an arXiv post \cite{ParrottCV2,ParrottCV3}.  In the former example our prescription for constructing contextual values in the case of a redundant detector (or underspecified measurement context) was not employed, and an anomalously divergent contextual value was inserted by hand; when our prescription for assigning contextual values is correctly applied, our theorem holds and a clear physical interpretation can be given to the measurement.  In the latter example a measurement context was chosen that cannot construct the desired observable to the lowest nonzero order in $g$, so our theorem does not apply.  Addressing these examples further demonstrates the power and utility of the contextual values formalism.

\ack
We acknowledge correspondence with S. Parrott.  This work was supported by the NSF Grant No. DMR-0844899, and ARO Grant No. W911NF-09-0-01417.


\end{document}